\documentstyle[12pt,oldlfont]{article}
\normalbaselines
\begin{document}
\bibliographystyle{unsrt}
\vbox {\vspace{6mm}}
\begin{center}
{\bf INTRODUCTION TO QUANTUM OPTICS}
\end{center}

\begin{center}
V. I. Man'ko
\end{center}

\bigskip

\begin{center}

Lebedev Physical Institute\\
Leninsky Prospekt, 53\\
Moscow 117924, Russia
\end{center}

\begin{abstract}

The theory of quantum propagator and time--dependent integrals of
motion in quantum optics is reviewed as well as the properties of
Wigner function, Q--function, and coherent state representation.
Propagators and wave functions of a free particle, harmonic oscillator,
and the oscillator with varying frequency are studied using
time--dependent linear in position and momentum integrals of motion.
Such nonclassical states of light (of quantum systems) as squeezed
states, correlated states, even and odd coherent states (Schr\"odinger
cat states) are considered. Photon distribution functions of
Schr\"odinger cat male and female states are given, and the photon
distribution function of squeezed vacuum is derived using the theory
of the oscillator with varying parameters. Properties of multivariable
Hermite polynomials used for the description of the multimode squeezed
and correlated light and polymode Schr\"odinger cats are studied.

\end{abstract}

\section{Introduction}
\label{s1}

In
the lectures the recent results \cite{dod89}, \cite{mal79} relating
time--dependent integrals of motion and Green function of quantum
system (Feynman path integral) are demonstrated. Obtaining of
the quantum propagator of a quantum system with quadratic Hamiltonians
is reduced to solving the system of the eigenvalue equations for the
linear in positions and momenta integrals of motion. In spite that in
the Schr\"odinger equation for the propagator the time and space
variables are not separated the existence of the time--dependent
invariants gives the possibility to treat this problem as the problem
with separable variables.

The integral of motion which is quadratic in position and momentum was
found for classical oscillator with time--dependent frequency a long
time ago by Ermakov \cite{Er1880}. Two time--dependent integrals of
motion which are linear forms in position and momentum for the classical
and quantum oscillator with time--dependent frequency were found in
\cite{mal70}; for a charge moving in varying in time uniform magnetic
field, this was done in \cite{MalTri69} (in the absense of the electric
field), \cite{MalJETP}, \cite{MalTriJETP} (nonstationary magnetic field
with the "circular" gauge of the vector potential plus uniform
nonstationary electric field, and \cite{MalDod72} (for the Landau gauge
of the time-dependent vector potential plus nonstationary electric field).
For the multimode nonstationary oscillatory systems, such new integrals
of motion, both of Ermakov's type (quadratic in positions and momenta)
and linear in position and momenta, generalizing the results of
\cite{mal70} were constructed in \cite{MalManTri73}. Recently
the discussed time--dependent invariants were obtained using
Noether's theorem in \cite{Cas}--\cite{Castanes}.

In the lectures the coherent states of the oscillator are discussed
and these states are used to construct Q--function of the oscillator
at zero and finite temperatures. The nonclassical states are discussed
in details. The Wigner function of multimode squeezed light is studied
using such special functions as multivariable Hermite polynomials. The
theory of parametric oscillator is appropriate to consider the problem
of creation of photons from vacuum in a resonator with moving
walls (with moving mirrors) which is the phenomenon based on the
existence of Casimir forces (so--called nonstationary Casimir effect).
The resonator with moving boundaries (moving mirrors, media with
time--dependent refractive index) produces also effect of squeezing
in the light quadratures. In the high energy physics very fast particle
collisions may produce new types of states of boson fields (pions, for
example) which are squeezed and correlated states studied in quantum
optics but almost unknown in particle physics, both theoretically and
experimentally.

\section{States of Quantum Systems}

In the Schr\"odinger representation, the state of one--dimensional
quantum system is described by a complex wave function (in coordinate
representation)
\begin{equation}
\psi(q,t)=<q|\psi ,t>,
\label{1}
\end{equation}
where $~|\psi ,t>$ is the time--dependent state vector in Hilbert
space of quantum system states, and $~|q>$ is the eigenstate of the
position operator. Such states are called pure states.

The mixed state of the quantum system is described by the density matrix
(a function of two variables $~q,q',$ and time $~t$)
\begin{equation}
\rho (q,q',t)=<q|\rho (t)|q'>,\label{2}
\end{equation}
where $~\rho (t)$ is the time--dependent density operator satisfying
the conditions of hermiticity
\begin{equation}
\rho \dag (t)=\rho (t),\label{3}
\end{equation}
normalization
\begin{equation}
\mbox {Tr}~\rho (t)=1,\label{4}
\end{equation}
and nonnegativity
\begin{equation}
\rho (t)\ge 0.\label{5}
\end{equation}
For pure state, the density operator which is projector on the state
$~|\psi ,t>$, i.e.
\begin{equation}
\rho (t)=|\psi ,t><\psi ,t|\label{6}
\end{equation}
satisfies the extra condition
\begin{equation}
\rho ^{2}(t)=\rho (t),\label{7}
\end{equation}
and, consequently,
\begin{equation}
\mu =\mbox {Tr} ~\rho ^{2}(t)=1.\label{8}
\end{equation}
For the system with a Hamiltonian operator $~H(t),$ the state vector
satisfies the Schr\"odinger equation ($~\hbar $ is Planck constant)
\begin{equation}
i\hbar \frac {\partial }{\partial t}|\psi ,t>=H(t)|\psi ,t>.\label{9}
\end{equation}
The density operator (\ref{6}) obeys the evolution equation
\begin{equation}
\frac {\partial \rho (t)}{\partial t}+\frac {i}{\hbar }[H(t),\rho (t)]
=0.\label{10}
\end{equation}
All the information on the quantum system is contained in the density
operator whose matrix elements may be taken in any appropriate basis.

\section{Evolution Operator}

The evolution of the state vector $~|\psi ,t>$ is determined by an
evolution operator $~U(t)$ which is defined by the equality
\begin{equation}
|\psi ,t>=U(t)|\psi ,0>.\label{11}
\end{equation}
This operator obeys the Schr\"odinger equation
\begin{equation}
i\hbar \frac {\partial U(t)}{\partial t}=H(t)U(t),\label{12}
\end{equation}
with the obvious initial condition
\begin{equation}
U(0)=\hat {\hbox {\bf 1}},\label{13}
\end{equation}
where $~\hat {\hbox {\bf 1}}$ is the unity operator. The matrix
element of the evolution operator in coordinate representation
is called Green function (or quantum propagator, or Feynman amplitude
considered as path integral)
\begin{equation}
G(q,q',t)=<q|U(t)|q'>.
\label{14}
\end{equation}
The Green function satisfies the Schr\"odinger equation
\begin{equation}
i\hbar \frac {\partial G(q,q',t)}{\partial t}-H(t)G(q,q',t)=
i\hbar \delta (t)\delta (q-q'),
\label{15}
\end{equation}
with the initial condition
\begin{equation}
G(q,q',0)=\delta (q-q').
\label{16}
\end{equation}
The propagator in momentum representation is connected with the
Green function by the integral transform
\begin{equation}
G(p,p',t)=<p|U(t)|p'>=\int <p|q><q|U(t)|q'><q'|p'>~dq~dq',
\label{17}
\end{equation}
which is the Fourier transform since
\begin{equation}
<q|p>=\frac {1}{\sqrt {2\pi \hbar }}\exp \frac {ipq}{\hbar },
\label{18}
\end{equation}
and we used the completeness condition of the position state vectors
\begin{equation}
\int |q><q|~dq=\hat {\hbox {\bf 1}}.
\label{19}
\end{equation}
Below we will use the propagator and density matrix in other
representations, too.

\section{Integrals of Motion}

The integral of motion or invariant operator $~I(t)$ is defined by
the equality
\begin{equation}
\frac {d}{dt}<\psi ,t|I(t)|\psi ,t>=0.
\label{20}
\end{equation}
The sufficient condition for the operator to be the invariant is
the equality to zero of the full time derivative of this operator,
\begin{equation}
\frac {dI(t)}{dt}=\frac {\partial I(t)}{\partial t}+\frac {i}{\hbar}
[H(t),I(t)]=0.
\label{21}
\end{equation}
Comparing Eq. (\ref{21}) with Eq. (\ref{10}) we see that the density
operator $~\rho (t)$ is the quantum integral of motion for systems
with Hamiltonian $~H(t)$.

There exist several theorems on the properties of integrals of
motion \cite{dod89}:\\
i) A function of integrals of motion $~F(I(t))$ is the integral
of motion.\\
ii) Eigenvalues of integrals of motion do not depend on time.\\
iii) Given a solution to Schr\"odinger equation $~|\psi (t)>$ and
integral of motion $~I(t)$. Then the vector $~|\phi ,t>=I(t)|\psi (t)>$
is solution to Schr\"odinger equation.

These properties are easily proved if one uses necessary and
sufficient conditions for the operator $~I(t)$ to be the integral of
motion which are expressed by the relation of the quantum invariant
and the evolution operator of the system $~U(t)$ of the form
\begin{equation}
I(t)=U(t)I(0)U^{-1}(t).
\label{22}
\end{equation}
Systems with hermitian Hamiltonians $~H\dag (t)=H(t)$ have
unitary evolution operators
\begin{equation}
U(t)\dag =U^{-1}(t).
\label{23}
\end{equation}
Observables $~I_{H}(t)$ in the Heisenberg representation are
related to the evolution operator by the formula
\begin{equation}
I_{H}(t)=U\dag (t)I(0)U(t).
\label{24}
\end{equation}
{}From Eqs. (\ref{22})--(\ref{24}), the relation of integrals of motion
$~I(t)$ to observables in the Heisenberg representation is
\cite{dod89}, \cite{mal79}
\begin{equation}
I(t)=U^{2}(t)I_{H}(t)U^{-2}(t).
\label{25}
\end{equation}
For stationary systems with time--independent Hamiltonians, this
relation is reduced to the equality
\begin{equation}
I(t)=I_{H}(-t).
\label{26}
\end{equation}
There are integrals of motion which at the initial time $~t=0$
coincide with the position and momentum operators. These invariants
are
\begin{equation}
p_{0}(t)=U(t)pU^{-1}(t),
\label{27}
\end{equation}
and
\begin{equation}
q_{0}(t)=U(t)qU^{-1}(t).
\label{28}
\end{equation}

\section{Invariants and Propagator}

{}From relations (\ref{27}), (\ref{28}) and the rule of
the multiplication of matrices, it follows the system of new
equations for the quantum propagator \cite{dod89}
\begin{equation}
p_{0}(t)_{q}G(q,q',t)=i\hbar \frac {\partial }{\partial q'}G(q,q',t),
\label{29}
\end{equation}
\begin{equation}
q_{0}(t)_{q}G(q,q',t)=q'G(q,q',t).
\label{30}
\end{equation}
In the left hand side of equalities subindices $~q$ mean that
operators $~p_{0}(t)$ and $~q_{0}(t)$ act on the argument $~q$ of
the Green function. The system of Eqs. (\ref{15}), (\ref{29}),
(\ref{30}) is selfconsistent system, and the coordinate dependence
of the quantum propagator may be found from Eqs. (\ref{29}),
(\ref{30}). The time--dependent phase factor of the Green function
is determined by Eq. (\ref{15}). Analogous systems of equations
for the quantum propagator may be written in any representation.

\section{Free Particle}

The Hamiltonian for a free particle with mass $~m$ is
time--independent
\begin{equation}
H=\frac {p^{2}}{2m}.
\label{31}
\end{equation}
The momentum operator is known to be the integral of motion for
the free motion. So we have
\begin{equation}
p_{0}(t)=p.
\label{32}
\end{equation}
It is easy to check that the time--dependent operator
\begin{equation}
q_{0}(t)=q-\frac {p}{m}t
\label{33}
\end{equation}
is the integral of motion since its full time derivative equals to zero.
Equality (\ref{26}) means that the position and momentum operators
of the free particle in the Heisenberg representation are
\begin{equation}
p_{H}(t)=p,
\label{34}
\end{equation}
\begin{equation}
q_{H}(t)=q+\frac {p}{m}t.
\label{35}
\end{equation}
If one introduces the vector--operator $~{\hbox {\bf Q}}$ with
components $~Q_{\alpha },~~\alpha =~1,~2,$ such that
$~Q_{1}=~p,$ $~Q_{2}=~q$, Eqs. (\ref{32}), (\ref{33}) may be written
in the matrix form
\begin{equation}
{\hbox {\bf Q}}_{0}(t)=\Lambda (t){\hbox {\bf Q}},
\label{36}
\end{equation}
where the 2$\times $2--matrix $~\Lambda (t)$ has the form
\begin{equation}
\Lambda (t)=\left (\begin{array}{clcr}
1&0\\
-t/m&1\end{array}\right ).
\label{37}
\end{equation}
One can check that the commutator
\begin{equation}
[Q_{\alpha },Q_{\beta }]=\Sigma _{\alpha \beta},
\label{38}
\end{equation}
where the matrix $~\Sigma $ has the form
\begin{equation}
\Sigma =-i\hbar \left (\begin{array}{clcr}
0&1\\
-1&0\end{array}\right ),
\label{39}
\end{equation}
is not changed for integrals of motion
\begin{equation}
[Q_{0\alpha }(t),Q_{0\beta }(t)]=\Sigma _{\alpha \beta}.
\label{40}
\end{equation}
It means that the matrix $~\Lambda (t)$ is the real symplectic
matrix, i.e.
\begin{equation}
\Lambda (t)\Sigma \Lambda ^{tr}(t)=\Sigma
\label{41}
\end{equation}
belonging to the Lie group $~Sp(2,R)$.

It is interesting to note that the time--dependent operator
quadratic in position and momentum
\begin{equation}
K(t)=q^{2}+\frac {p^{2}t^{2}}{m^{2}}-\frac {t}{m}(pq+qp)
\label{42}
\end{equation}
is the integral of motion for the free particle. This operator
is obtained using the relation
\begin{equation}
K(t)=q_{0}^{2}(t)
\label{43}
\end{equation}
analogous to the relation of the energy to the invariant momentum
$~p_{0}(t)$ (\ref{32})
$$H=\frac {1}{2m}p_{0}^{2}(t).$$
According to Eqs. (\ref{29}), (\ref{30}) and (\ref{32}), (\ref{33})
the free particle propagator must satisfy the system of equations
\begin{equation}
-i\hbar \frac {\partial G(q,q',t)}{\partial q}
=i\hbar \frac {\partial G(q,q',t)}{\partial q'},
\label{44}
\end{equation}
\begin{equation}
qG(q,q',t)
+\frac {i\hbar t}{m}\frac {\partial G(q,q',t)}{\partial q}
=q'G(q,q',t).
\label{45}
\end{equation}
Equation (\ref{44}) means that
\begin{equation}
G(q,q',t)=g(x,t),~~~~x=q-q',
\label{46}
\end{equation}
where $~g(x,~t)$ is an arbitrary function.
Equation (\ref{45}) may be rewritten in the form
\begin{equation}
xg(x,t)+\frac {i\hbar t}{m}\frac {\partial g(x,t)}{\partial x}
=0.
\label{47}
\end{equation}
Then we obtain the solution for this equation in Gaussian form
\begin{equation}
g(x,t)=N(t)\exp \frac {imx^{2}}{2\hbar t}.
\label{48}
\end{equation}
The function $~N(t)$ is obtained from Eq. (\ref{15})
\begin{equation}
N(t)=\frac {C}{\sqrt t}.
\label{49}
\end{equation}
The normalization constant $~C$ can be found from the initial
condition for the propagator (\ref{16}) using the formula
\begin{equation}
\lim_{\varepsilon \rightarrow 0}\{\frac {1}{\sqrt {\varepsilon \pi }}
\exp -\frac {x^{2}}{\varepsilon }\}=\delta (x),
\label{50}
\end{equation}
which gives the free particle propagator
\begin{equation}
G(q,q',t)={\sqrt \frac {m}{2\pi i\hbar t}}\exp \frac {im(q
-q')^{2}}{2\hbar t}.
\label{51}
\end{equation}

\section{Harmonic Oscillator}

Let us study by methods of the integrals of motion the harmonic
oscillator with the time--independent Hamiltonian
\begin{equation}
H=\frac {p^{2}}{2m}+\frac {m\omega ^{2}q^{2}}{2}.
\label{52}
\end{equation}
One can check that there exist two invariants linear in position
and momentum
\begin{equation}
p_{0}(t)=\cos \omega t~~p+m\omega \sin \omega t~~q
\label{53}
\end{equation}
and
\begin{equation}
q_{0}(t)=-\frac {1}{m\omega }\sin \omega t~~p+\cos \omega t~~q.
\label{54}
\end{equation}
For $~\omega \rightarrow 0$ invariants (\ref{53}), (\ref{54})
coincide with free motion invariants (\ref{32}), (\ref{33}),
respectively.

The propagator of the harmonic oscillator satisfies the system of
equations
\begin{equation}
-i\hbar \cos \omega t~~\frac {\partial G(q,q',t)}{\partial q}
+m\omega q\sin \omega t~~G(q,q',t)
=i\hbar \frac {\partial G(q,q',t)}{\partial q'},
\label{55}
\end{equation}
\begin{equation}
\frac {i\hbar }{m\omega }\sin \omega t~~\frac {\partial G(q,q',t)}
{\partial q}+q\cos \omega t~~G(q,q',t)=q'G(q,q',t),
\label{56}
\end{equation}
\begin{equation}
i\hbar \frac {\partial G(q,q',t)}{\partial t}=
-\frac {\hbar ^{2}}{2m}\frac {\partial ^{2}G(q,q',t)}{\partial ^{2}q}
+\frac {m\omega ^{2}}{2}q^{2}G(q,q',t),~~~t>0.
\label{57}
\end{equation}
Integrating Eq. (\ref{56}) we find $~q$--dependence of the propagator
\begin{equation}
G(q,q',t)=f(q',t)\exp \{\frac {im\omega }{2\hbar }(q^{2}\cot \omega t
{}~~-\frac {2qq'}{\sin \omega t})\},
\label{58}
\end{equation}
in which the factor $~f(q',~t)$ has to be found using Eqs. (\ref{55})
and (\ref{57}).

Using Eq. (\ref{55}) we obtain the following relation
\begin{equation}
\frac {\partial f(q',t)}{\partial q'}=q'\frac {im\omega }{\hbar }
\cot \omega t~~f(q',t).
\label{59}
\end{equation}
{}From this equation $~q'$--dependence of the propagator is found and
we have
\begin{equation}
G(q,q',t)=N(t)\exp \{\frac {im\omega }{2\hbar }[(q^{2}
+q'^{2})\cot \omega t~~-\frac {2qq'}{\sin \omega t}]\}.
\label{60}
\end{equation}
Due to Eq. (\ref{57}) the time--dependent factor $~N(~t)$ satisfies
the equation
\begin{equation}
\dot N(t)=-\frac {\omega }{2}\cot {\omega }{t}~~N(t),
\label{61}
\end{equation}
which gives
\begin{equation}
N(t)=\frac {C}{\sqrt {\sin \omega t}}.
\label{62}
\end{equation}
The normalization constant $~C$ may be found from the initial condition
(\ref{16}) using Eq. (\ref{50}). So we have the propagator of the
harmonic oscillator
\begin{equation}
G(q,q',t)={\sqrt \frac {m\omega }{2\pi i\hbar \sin \omega t}}
\exp \{\frac {im\omega }{2\hbar }[(q^{2}
+q'^{2})\cot \omega t~~-\frac {2qq'}{\sin \omega t}]\}.
\label{63}
\end{equation}
Equation (\ref{57}) for the propagator of the oscillator
is the equation in which variables $~q,~q',~t$ are not
separable. The advantage of the metod of integrals of motion is
the possibility to use the procedure analogous to the procedure
when we solve the equation with separable variables. Thus we find
the dependence of the propagator on different coordinates step by
step as in the method of separable coordinates.

\section{Coherent States}

Let us introduce the annihilation and creation operators
\begin{equation}
a=\frac {1}{\sqrt 2}(\sqrt {\frac {m\omega }{\hbar }}q+\frac {i}
{\sqrt {\hbar m\omega }}p),
\label{64}
\end{equation}
\begin{equation}
a\dag =\frac {1}{\sqrt 2}(\sqrt {\frac {m\omega }{\hbar }}q-\frac {i}
{\sqrt {\hbar m\omega }}p).
\label{65}
\end{equation}
They satisfy commutation relations
\begin{equation}
[a,a\dag ]=\hat {\hbox {\bf 1}}.
\label{66}
\end{equation}
The Hamiltonian of the harmonic oscillator can be rewritten as
\begin{equation}
H=\hbar \omega (a\dag a+\frac {1}{2}).
\label{67}
\end{equation}
The integrals of motion linear in the creation and annihilation
operators are \cite {dod89}, \cite {mal79}, \cite {mal70}
\begin{equation}
A(t)=e^{i\omega t}a,
\label{68}
\end{equation}
\begin{equation}
A\dag (t)=e^{-i\omega t}a\dag .
\label{69}
\end{equation}
The operator $~A(t)$ is the linear combination of invariants
$~p_{0}(t)$ (\ref{53}) and $~q_{0}(t)$ (\ref{54}) of the form
\begin{equation}
A(t)=\frac {1}{\sqrt 2}[\sqrt {\frac {m\omega }{\hbar }}q_{0}(t)
+\frac {i}{\sqrt {\hbar m\omega }}p_{0}(t)].
\label{70}
\end{equation}
Let us find normalized solutions to the eigenvalue equation
\begin{equation}
a\psi _{\alpha }(q)=\alpha \psi _{\alpha }(q),
\label{71}
\end{equation}
where $~\alpha $ is a complex number. Equation (\ref{71})
rewritten as the differential equation
\begin{equation}
\sqrt {\frac {m\omega }{\hbar }}q\psi _{\alpha }(q)+\frac {1}
{\sqrt {\hbar m\omega }}\frac {\partial \psi _{\alpha }(q)}
{\partial q}=\sqrt {2}\alpha \psi _{\alpha }(q)
\label{72}
\end{equation}
has the normalized solution of the Gaussian form
\begin{equation}
\psi _{\alpha }(q)=(\frac {m\omega }{\hbar \pi })^{1/4}
\exp (-\frac {|\alpha |^{2}}{2}-\frac {m\omega }{2\hbar }q^{2}
+\sqrt {\frac {2m\omega }{\hbar }}\alpha q-\frac {\alpha ^{2}}{2}).
\label{73}
\end{equation}
For $~\alpha =0,$ we have the wave function of the ground state
\begin{equation}
\psi _{0}(q)=<q|0>=(\frac {m\omega }{\hbar \pi })^{1/4}
\exp (-\frac {m\omega }{2\hbar }q^{2}).
\label{74}
\end{equation}
The state $~|\alpha >$ which is the normalized eigenstate of the
annihilation operator $~a$
\begin{equation}
a|\alpha >=\alpha |\alpha >
\label{75}
\end{equation}
is called the coherent state and in coordinate representation
the wave function of this state is given by Eq. (\ref{73}).
Orthogonal number states $~|n>,~~n=~0,~1,~2,\ldots $ which are
eigenstates of the hermitian operator $~a\dag a$,
\begin{equation}
a\dag a|n>=n|n>,
\label{76}
\end{equation}
satisfy the completeness condition
\begin{equation}
\sum _{n=0}^{\infty }|n><n|=\hat {\hbox {\bf 1}}.
\label{77}
\end{equation}
The coherent state is decomposed as
\begin{equation}
|\alpha >=\exp -\frac {|\alpha |^{2}}{2}~\sum _{n=0}^{\infty }
\frac {\alpha ^{n}}{\sqrt {n!}}|n>.
\label{78}
\end{equation}
Since
\begin{eqnarray}
a|n>&=&\sqrt {n}|n-1>,\nonumber\\
a\dag |n>&=&\sqrt {n+1}|n+1>,\label{79}
\end{eqnarray}
one can check that the series (\ref{78}) satisfies the eigenvalue
equation (\ref{75}). Due to the orthogonality property of the number
state basis
\begin{equation}
<n|m>=\delta _{nm},
\label{80}
\end{equation}
one has the scalar product of coherent states or matrix
elements of unity operator in the coherent state basis
\begin{equation}
<\alpha |\beta >=\exp (-\frac {|\alpha |^{2}}{2}-\frac {|\beta |}{2}
+\alpha ^{*}\beta )=<\alpha |\hat {\hbox {\bf 1}}|\beta >.
\label{81}
\end{equation}
Using the integral
\begin{equation}
\int _{-\infty }^{\infty }\exp (-ax^{2}+bx)~dx=\sqrt {\frac {\pi }{a}}
\exp \frac {b^{2}}{4a},
\label{82}
\end{equation}
and differentiating this equality as the function of variable
$~a$ we have
\begin{equation}
\int _{-\infty }^{\infty }x^{2}\exp (-ax^{2}+bx)~dx
=\frac {\sqrt {\pi }}{2}a^{-3/2}(1
+\frac {b^{2}}{2a})\exp \frac {b^{2}}{4}.
\label{83}
\end{equation}
Differentiating equality (\ref{82}) as the function of
variable $~b$ we have
\begin{equation}
\int _{-\infty }^{\infty }x\exp (-ax^{2}+bx)~dx
=\frac {b}{2a}\sqrt {\frac {\pi }{a}}\exp \frac {b^{2}}{4a}.
\label{84}
\end{equation}
The mean value of the oscillator coordinate in the coherent state
may be calculated using the integral (\ref{84})
\begin{equation}
<\alpha |q|\alpha >=\int _{-\infty }^{\infty }q|\psi _{\alpha }
(q)|^{2}~dq=\sqrt {\frac {2\hbar }{m\omega }}~\mbox {Re}~\alpha .
\label{85}
\end{equation}
The mean value of momentum in the coherent state $~|\alpha >$
is also given by the integral of the form (\ref{84}), and we have
\begin{equation}
<\alpha |p|\alpha >
=-i\hbar \int _{-\infty }^{\infty }\psi _{\alpha }^{*}(q)
\frac {\partial \psi _{\alpha }(q)}{\partial q}~dq
=\sqrt {2\hbar m\omega }~\mbox {Im}~\alpha .
\label{86}
\end{equation}
The dispersion of the coordinate in the coherent state
\begin{equation}
\sigma _{q}=<\alpha |q^{2}|\alpha >-(<\alpha |q|\alpha >)^{2}
\label{87}
\end{equation}
may be calculated using the integral (\ref{83}) and one obtaines
\begin{equation}
\sigma _{q}=\frac {\hbar }{2m\omega }.
\label{88}
\end{equation}
The dispersion of the momentum in the coherent state
\begin{equation}
\sigma _{p}=<\alpha |p^{2}|\alpha >-(<\alpha |p|\alpha >)^{2}
\label{89}
\end{equation}
may be also obtained by means of the integral (\ref{83}) and one has
\begin{equation}
\sigma _{p}=\frac {\hbar m\omega }{2}.
\label{90}
\end{equation}
Thus dispersions of momentum and position which in quantum
optics are called dispersions of quadratures do not depend on
the complex number $~\alpha $. The covariance of the position and
momentum in the coherent state
\begin{equation}
\sigma _{qp}=\frac {1}{2}<\alpha |(qp+pq)|\alpha >
-<\alpha |q|\alpha ><\alpha |p|\alpha >
\label{91}
\end{equation}
turns out to be equal to zero
\begin{equation}
\sigma _{qp}=0.
\label{92}
\end{equation}
The decomposition of the coherent state (\ref{78}) means that the
photon distribution function of the coherent light
\begin{equation}
P_{n}=|<n|\alpha >|^{2}=e^{-|\alpha |^{2}}\frac {|\alpha |^{2n}}{n!}
\label{93}
\end{equation}
is the Poisson distribution function. Thus in coherent states
uncertainties of quadratures are equal to uncertainties
in the vacuum state $~|0>$, and dispersions minimize
the Heisenberg uncertainty relation
\begin{equation}
\sigma _{p}\sigma _{q}\ge \frac {\hbar ^{2}}{4},
\label{94}
\end{equation}
which for coherent state becomes the equality
\begin{equation}
\sigma _{p}\sigma _{q}=\frac {\hbar ^{2}}{4}.
\label{95}
\end{equation}
There are no quadrature correlations in the coherent state. The mean
value of the number of photons in the coherent state is
\begin{equation}
<n>=\sum _{n=0}^{\infty }nP(n)=|\alpha |^{2},
\label{96}
\end{equation}
and the photon number dispersion equals
\begin{equation}
\sigma _{n}=<n^{2}>-<n>^{2}=|\alpha |^{2},
\label{97}
\end{equation}
since for the Poisson distribution the variance is equal to mean
value of a random variable. Using the generating function for the Poisson
distribution
\begin{equation}
G(z)=\sum _{n=0}^{\infty }z^{n}P(n)=\exp [(z-1)|\alpha |^{2}],
\label{98}
\end{equation}
one can easily check these properties due to relations
\begin{eqnarray}
\frac {dG}{dz}(z=1)&=&<n>,\nonumber\\
\frac {d^{2}G}{d^{2}z}(z=1)&=&<n>^{2}-<n>.
\label{99}
\end{eqnarray}
The coherent state may be created from the vacuum state by means
of the unitary displacement operator
\begin{equation}
|\alpha >=D(\alpha )|0>,
\label{100}
\end{equation}
where
\begin{equation}
D(\alpha )=\exp (\alpha a\dag -\alpha ^{*}a)
=\exp (-\frac {|\alpha |^{2}}{2})~\exp (\alpha a\dag )
{}~\exp (-\alpha ^{*}a).
\label{101}
\end{equation}
For matrices $~A$ and $~B$ of finite and infinite dimensions, there
exists the formula
\begin{equation}
e^{B}Ae^{-B}=A+[B,A]+\frac {1}{2!}[B,[B,A]]
+\frac {1}{3!}[B,[B,[B,A]]]+\ldots
\label{102}
\end{equation}
In the case
$$[A,[A,B]]=[B,[A,B]]=0,$$
the following rule holds
\begin{equation}
e^{A}~e^{B}=\exp (A+B+\frac {1}{2}[A,B]).
\label{103}
\end{equation}
This rule was used in formula (\ref{101}) since the commutator
of the creation and annihilation operators commutes with these
operators. Applying the displacement operator to vacuum state and
using the equality
\begin{equation}
\exp (-\alpha ^{*}a)|0>=|0>
\label{104}
\end{equation}
we reproduce the series (\ref{78}) since the number state $~|n>$ is
given by the relation
\begin{equation}
|n>=\frac {a\dag ^{n}}{\sqrt {n!}}|0>.
\label{105}
\end{equation}
{}From the decomposition (\ref{78}) it is obvious that the vector
$~\exp (|\alpha |^{2}/2)~|\alpha >$ is analytical function of the
complex variable $~\alpha $, and it is the generating function for
the number state vector $~|n>$. Using this remark and the definition
of Hermite polynomials through the generating function
\begin{equation}
e^{-\alpha ^{2}+2t\alpha }=\sum _{n=0}^{\infty }\frac {H_{n}(t)}{n!}
\alpha ^{n},
\label{106}
\end{equation}
one can obtain the expression for the wave function of the number
state $~\psi _{n}(q)=~<q|n>$ comparing the coherent state wave
function (\ref{73}) with the generating function (\ref{106}). We
obtain
\begin{equation}
\psi _{n}(q)=\psi _{0}(q)\frac {1}{\sqrt {2^{n}n!}}
H_{n}(q\sqrt {\frac {m\omega }{\hbar }}),
\label{107}
\end{equation}
where $~\psi _{0}(q)$ is given by Eq. (\ref{74}).

Using the completeness relation of number states (\ref{77})
one can obtain the completeness relation of coherent states
\begin{equation}
\frac {1}{\pi }\int \int _{-\infty }^{\infty }|\alpha ><\alpha |
{}~d\alpha _{1}~d\alpha _{2}=\hat {\hbox {\bf 1}},~~~~~\alpha =
\alpha _{1}+i\alpha _{2},
\label{108}
\end{equation}
since in polar coordinates the integral in the left hand side
of equality (\ref{108}) is reduced easily to the sum in
the left hand side of Eq. (\ref{77}).

\section{Evolution of Coherent State}

Since the Green function is the kernel of the evolution operator
the evolution of the coherent state wave function is given by the
integral
\begin{equation}
\psi _{\alpha }(q,t)=\int _{-\infty }^{\infty }G(q,q',t)
\psi _{\alpha }(q')~dq',
\label{109}
\end{equation}
where we use the propagator (\ref{63}). The corresponding integral
is of the Gaussian form (\ref{82}) and we have
\begin{equation}
\psi _{\alpha }(q,t)=e^{-i\omega t/2}\psi _{\widetilde \alpha }
(q),~~~~~\widetilde \alpha =\alpha e^{-i\omega t }.
\label{110}
\end{equation}
Thus in the process of evolution the initial coherent state of the
harmonic oscillator preserves the property to be the eigenvector of
the annihilation operator, but its eigenvalue has the time--dependent
phase factor corresponding to the classical motion of the oscillator
in its phase space due to formulae (\ref{85}), (\ref{86}). It means
that in the process of evolution dispersions of quadratures
do not depend on time and no quadrature correlation appears
due to the evolution. Having this result it is easy to obtain the
oscillator propagator in the coherent state basis. We have by definition
\begin{eqnarray}
G(\alpha ^{*},\beta ,t)=<\alpha |U(t)|\beta >&=&e^{-i\omega t/2}
<\alpha |\beta e^{-i\omega t}>\nonumber\\
&=&e^{-i\omega t/2}\exp [-\frac {|\alpha |^{2}}{2}-
\frac {|\beta |^{2}}{2}+\alpha ^{*}\beta e^{-i\omega t}].
\label{111}
\end{eqnarray}
Using the property of coherent states to be the generating function
for number states and expanding the function
$~\exp (\frac {|\alpha |^{2}}{2}+\frac {|\beta |^{2}}{2})~G(~\alpha ^{*},
{}~\beta ,~t)$ into the power series in variables $~\alpha ^{*},~\beta $
we obtain the obvious expression for the oscillator propagator in the
Fock basis
\begin{equation}
G(n,m,t)=\delta _{nm}e^{-i\omega t(n+1/2)}.
\label{112}
\end{equation}

\section{Squeezing in Parametric Oscillator}

For the parametric oscillator with the Hamiltonian
\begin{equation}
H=-\frac {\partial ^{2}}{2\partial x^{2}}
+\frac {\omega ^{2}(t)x^{2}}{2},
\label{s60}
\end{equation}
where we take $~\hbar=~m=~\omega (0)=~1$, there exists the
time--dependent integral of motion found in \cite {mal70}
\begin{equation}
A=\frac {i}{\sqrt 2}[\varepsilon (t)p-\dot \varepsilon (t)x],
\label{s61}
\end{equation}
where
\begin{equation}
\ddot \varepsilon (t)+\omega ^{2}(t)\varepsilon (t)=0,~~~~~~
\varepsilon (0)=1,~~~~~~\dot \varepsilon (0)=i,
\label{s62}
\end{equation}
satisfying the commutation relation
\begin{equation}
[A,~A\dag ]=1.
\label{s63}
\end{equation}
It is easy to show that packet solutions of the Schr\"odinger equation
may be introduced and interpreted as coherent states \cite{mal70}, since
they are eigenstates of the operator $~A$ (\ref{s61}), of the form
\begin{equation}
\Psi _{\alpha }(x,t)=\Psi _{0}(x,t)\exp \{-\frac {|\alpha |^{2}}{2}-
\frac {\alpha ^{2}\varepsilon ^{*}(t)}{2\varepsilon (t)}
+\frac {{\sqrt 2}\alpha x}{\varepsilon}\},
\label{s64}
\end{equation}
where
\begin{equation}
\Psi _{0}(x,t)=\pi ^{-1/4}\varepsilon (t)^{-1/2}
\exp \frac {i\dot \varepsilon (t)x^{2}}{2\varepsilon (t)}
\label{s65}
\end{equation}
is analog of the ground state of the oscillator and $~\alpha $ is a
complex number. Variances of the position and momentum of the
parametric oscillator in the state (\ref{s65}) are
\begin{equation}
\sigma _{x}=\frac {|\varepsilon (t)|^{2}}{2},~~~~~~\sigma _{p}
=\frac {|\dot \varepsilon (t)|^{2}}{2},
\label{s66}
\end{equation}
and the correlation coefficient $~r$ of the position and momentum has
the value corresponding to minimization of the Schr\"odinger uncertainty
relation \cite{schrod}
\begin{equation}
\sigma _{x}\sigma_{p}=\frac {1}{4}\frac {1}{1-r^{2}}.
\label{s67}
\end{equation}
If $~\sigma _{x}<1/2~~(\sigma _{p}<1/2)$ we have squeezing in photon
quadrature components.

The analogs of orthogonal and complete system of states which are excited
states of stationary oscillator are obtained by expansion of (\ref{s64})
into power series  in $~\alpha .$ We have
\begin{equation}\label{insert1}
\Psi _{m}(x,t)=\left (\frac {\varepsilon ^{*}(t)}{2\varepsilon (t)}
\right )^{m/2}\frac {1}{\sqrt {m!}}\Psi _{0}(x,t)H_{m}\left (\frac {x}
{|\varepsilon (t)|}\right ),
\end{equation}
and these squeezed and correlated number states are eigenstates of
invariant $~A^{\dag }A.$

The function (\ref{s65}) describes the squeezed vacuum state. The photon
distribution function in the squeezed vacuum is expressed in terms
of the overlap integral
\begin{equation}
W(n)=|C(n)|^{2}
=|\int_{\infty }^{\infty }\psi _{0}^{*}(x,t)\psi _{n}(x)~dx|^{2},
\label{d1}
\end{equation}
where the function $~\psi _{n}(x)$ is the wave function of the $~n$--photon
state (\ref{107}). The amplitude $~C(n)$ can be calculated using the
overlap integral
\begin{equation}
\phi (\alpha )
=\int_{\infty }^{\infty }\psi _{0}^{*}(x,t)e^{|\alpha |^{2}/2}
\psi _{\alpha }(x)~dx,
\label{d2}
\end{equation}
which is generating function for the amplitudes $~C(n)$
\begin{equation}
\phi (\alpha )=\sum _{n=0}^{\infty }\frac {\alpha ^{n}}{\sqrt {n!}}C_{n},
\label{d3}
\end{equation}
since according to Eq. (\ref{78}) the wave function of the coherent state
$~e^{|\alpha |^{2}/2}\psi _{\alpha }(x)$ (\ref{73}) is the generating
function for the $~n$--photon state. The Gaussian integral (\ref{d2})
is (we have $~\hbar =~m=~\omega =~1)$
\begin{equation}
\phi _{\alpha }=\int _{-\infty }^{\infty }
\frac {dx}{\sqrt {\pi \varepsilon ^{*}}}\exp [-\frac {\alpha ^{2}}{2}-
\frac {1}{2}(1+\frac {i\dot \varepsilon ^{*}}{\varepsilon ^{*}})x^{2}+
\sqrt {2}\alpha x],
\label{d4}
\end{equation}
and using formula (\ref{82}) we obtain
\begin{equation}
\phi _{\alpha }=C(0)\exp \alpha ^{2}\mu,
\label{d5}
\end{equation}
where
\begin{equation}
C(0)=\sqrt {\frac {2}{\varepsilon ^{*}+i\dot \varepsilon ^{*}}},~~~~~
\mu=\frac {\varepsilon ^{*}-i\dot \varepsilon ^{*}}{2(\varepsilon ^{*}
+i\dot \varepsilon ^{*})}.
\label{d6}
\end{equation}
The probability to have no photons is
\begin{equation}
W(0)=\frac {2}{\sqrt {|\varepsilon |^{2}+|\dot \varepsilon |^{2}+2}}.
\label{d7}
\end{equation}
Expanding Eq. (\ref{d4}) into power series in $~\alpha $ we obtain the
result that the probability to have the odd number of photons $~n=~2m+1$
is equal to zero
\begin{equation}
W(2m+1)=0,
\label{d8}
\end{equation}
and the probability to have even number of photons $~n=~2m$ equals
\begin{equation}
W(2m)=W(0)\frac {(2m)!}{(m!)^{2}}|\mu |^{2m}.
\label{d9}
\end{equation}
The partial cases of parametric oscillator are free motion
$~(~\omega (t)=0~),$ starionary harmonic oscillator
$~(~\omega ^{2}(t)=1~),$ and repulsive oscillator
$~(~\omega ^{2}(t)=-1~).$ The obtained above solutions are described
by the function $~\varepsilon (t)~$ which is equal to
$~\varepsilon (t)=1+it,$ for free particle, $~\varepsilon (t)=e^{it},$
for usual oscillator, and $~\varepsilon (t)=\cosh t+i\sinh t,$ for
repulsive oscillator.

If one introduces the squeezing parameter $~r$ (for zero correlation)
according to
\begin{equation}
|\varepsilon |=e^{-r},~~~~~|\dot \varepsilon |=e^{r},
\label{d10}
\end{equation}
the distribution (\ref{d9}) may be rewritten as
\begin{equation}
W(2m)=\frac {1}{\cosh r}\frac {(2m)!}{(m!)^{2}}(\frac {\tanh r}{2})^{2m}.
\label{d11}
\end{equation}

Another normalized solution to the Schr\"odinger equation
\begin{equation}
\Psi _{\alpha m}(x,t)=2N_{m}\Psi _{0}(x,t)\exp \{-\frac {|\alpha |^{2}}
{2}-\frac {\varepsilon ^{*}(t)\alpha ^{2}}{2\varepsilon (t)}\}\cosh
\frac {{\sqrt 2}\alpha x}{\varepsilon (t)},
\label{s68}
\end{equation}
where
\begin{equation}
N_{m}=\frac {\exp (|\alpha |^{2}/2)}{2\sqrt {\cosh |\alpha |^{2}}},
\label{s69}
\end{equation}
is the even coherent state \cite{dod74} (the Schr\"odinger cat male
state). The odd coherent state of the parametric oscillator
(Schr\"odinger cat female state)
\begin{equation}
\Psi _{\alpha f}(x,t)=2N_{f}\Psi _{0}(x,t)\exp \{-\frac {|\alpha |^{2}}
{2}-\frac {\varepsilon ^{*}(t)\alpha ^{2}}{2\varepsilon (t)}\}\sinh
\frac {\sqrt {2}\alpha x}{\varepsilon (t)},
\label{s70}
\end{equation}
where
\begin{equation}
N_{f}=\frac {\exp (|\alpha |^{2}/2)}{2\sqrt {\sinh |\alpha |^{2}}},
\label{s71}
\end{equation}
satisfies the Schr\"odinger equation and is the eigenstate of the
integral of motion $~A^{2}$ (as well as the even coherent state)
with the eigenvalue $~\alpha ^{2}$. These states are one--mode examples
of squeezed and correlated Schr\"odinger cat states constructed in
\cite{nikon}.

\section{Wigner and Q--functions}

Let us discuss now other possible descriptions of the quantum state.
There exist two important functions which we describe for multimode
system with $~N$ degrees of freedom.
The Wigner function of a system  $~W({\bf p,q})=~W({\bf Q})$ is
expressed in terms of density matrix in dimensionless coordinate
representation as (see, \cite{wigner32})
\begin{equation}
W({\bf p,q})=\int \rho ({\bf q}+\frac {\bf u}{2},~{\bf q}
-\frac {\bf u}{2})\exp (-i{\bf p}{\bf u})~d{\bf u}.
\label{w44}
\end{equation}
The inverse transform is
\begin{equation}
\rho ({\bf x,x'})=\frac {1}{(2\pi )^{N}}\int W(\frac {{\bf x}
+{\bf x'}}{2},~{\bf p})\exp [i{\bf p}({\bf x}-{\bf x'})]~d{\bf p}.
\label{w45}
\end{equation}
The Q--function \cite{hus40} is expressed in terms of Wigner function
through the 3N--dimensional integral transform
\begin{equation}
Q({\bf B})
=\frac {1}{(2\pi )^{N}}\int \Phi _{\bf B}({\bf x,x',p})W(\frac
{{\bf x}+{\bf x'}}{2},{\bf p})~d{\bf x}~d{\bf x'}~d{\bf p},
\label{w46}
\end{equation}
with the kernel
\begin{equation}
\Phi _{\bf B}({\bf x,x',p})=\pi ^{-N/2}\exp [i{\bf p}({\bf x}
-{\bf x'})-\frac {1}{2}{\bf B}(\sigma _{Nx}+{\bf I}_{2N}){\bf B}
-\frac {{\bf X}^{2}}{2}+{\sqrt 2}{\bf B}\sigma _{Nx}{\bf X}],
\label{w47}
\end{equation}
where the 2N--vector $~{\bf X}=({\bf x,x'})$ is introduced. If one
has the Q--function the Wigner function is given by the integral
transform
\begin{equation}
W({\bf p,q})=\frac {1}{\pi ^{2N}}\int \{\prod _{k=1}^{N}d^{2}
\beta _{k}~d^{2}\gamma _{k}~du_{k}\widetilde \Phi _{k}(u_{k},
\widetilde {\bf B})\}
Q(\widetilde {\bf B}),
\label{w48}
\end{equation}
where the argument of the Q--function $~{\bf B}$ is replaced by the
2N--vector with complex components
$$\widetilde {\bf B}=(~\beta _{1},~\beta _{2},~\ldots ,~\beta _{N},
{}~\gamma _{1}^{*},~\gamma _{2}^{*},~\ldots ,~\gamma _{N}^{*}),$$
and the kernel has the form
\begin{eqnarray}
\widetilde \Phi _{k}(u_{k},\widetilde {\bf B})&=&\pi ^{-1/2}
\exp [-|\beta _{\dot k}|^{2}
-|\gamma _{\dot k}|^{2}+{\sqrt 2}(q_{\dot k}
+\frac {u_{\dot k}}{2})\beta _{\dot k}+{\sqrt 2}(q_{\dot k}
-\frac {u_{\dot k}}{2})\gamma _{\dot k}^{*}\nonumber\\
&-&\frac {1}{2}(q_{k}+\frac {u_{k}}{2})^{2}-\frac {1}{2}(q_{k}
-\frac {u_{k}}{2})^{2}-\frac {\beta _{\dot k}^{2}}{2}
-\frac {\gamma _{\dot k}^{*2}}{2}
-ip_{\dot k}u_{\dot k}+\gamma _{\dot k}\beta _{\dot k}^{*}].
\label{w49}
\end{eqnarray}
The density matrix in coordinate representation is related to the
Q--function
\begin{equation}
\rho ({\bf x,x'})=\pi ^{-2N}\int \{\prod _{k=1}^{N}d^{2}\beta _{k}
{}~d^{2}\gamma _{k}\phi_{k} (\widetilde {\bf B})
\exp [-\frac {1}{2}(x_{k}^{2}+x_{k}^{'2})]\}
Q(\widetilde {\bf B}),
\label{w50}
\end{equation}
where the kernel of the transform is
\begin{equation}
\phi _{k}(\widetilde {\bf B})=\pi ^{-1/2}
\exp [-|\beta _{k}|^{2}
-|\gamma _{k}|^{2}+{\sqrt 2}x_{k}\beta _{k}
+{\sqrt 2}x'_{k}\gamma _{k}^{*}
-\frac {\beta _{k}^{2}}{2}-\frac {\gamma _{k}^{*2}}{2}
+\gamma _{k}\beta _{k}^{*}].
\label{w51}
\end{equation}
The evolution of the Wigner function and Q--function for systems
with quadratic Hamiltonians for any state is given by
the following prescription. Given the Wigner function $~W({\bf p,q},t=0)$
for the initial moment of time $~t=0.$ Then the Wigner function for time
$~t$ is obtained by the replacement
$$W({\bf p,q},t)=~W({\bf p}(t),~{\bf q}(t),~t=0),$$
where time--dependent arguments are linear integrals of motion
of the quadratic system found in \cite{dod89}, \cite{MalManTri73} and
\cite{tri73}. The same
ansatz is used for the Q--function. Namely, given the Q--function
of the quadratic system $~Q({\bf B},~(t=0))$ for the initial moment of
time $~t=0.$ Then the Q--function for time $~t$ is given by
the replacement
$$Q({\bf B},~t)=~Q({\bf B}(t),~t=0),$$
where the 2N--vector $~{\bf B}(t)$ is the integral of motion linear in the
annihilation and creation operators found in \cite{dod89},
\cite{MalManTri73} and \cite{tri73}.
This ansatz follows from the statement that the density operator of the
Hamiltonian system is the integral of motion, and its matrix elements in
any basis must depend on appropriate integrals of motion. In
particular, the Wigner function and Q--function depend just on linear
invariants found in \cite{dod89}, \cite{MalManTri73} and \cite{tri73}.

As an example we consider the Wigner function of the oscillator ground
state with the density matrix in coordinate representation (we
reconstruct below the dimensionality of variables)
\begin{equation}
\rho _{0}(x,x')
=\sqrt {\frac {m\omega }{\hbar \pi }}\exp [-\frac {1}{2}
\frac {m\omega }{\hbar }(x^{2}+x'^{2})].
\label{v1}
\end{equation}
Using the definition of the Wigner function we have
\begin{equation}
W_{0}(p,q)
=\sqrt {\frac {m\omega }{\hbar \pi }}\int _{-\infty}^{\infty }
\exp \{-\frac {1}{2}\frac {m\omega }{\hbar }[(q-\frac {u}{2})^{2}
+(q+\frac {u}{2})^{2}]-\frac {ipu}{\hbar }\}~du,
\label{v2}
\end{equation}
and calculating the Gaussian integral we obtain
\begin{equation}
W_{0}(p,q)=2\exp (-\frac {p^{2}}{\hbar m \omega }
-\frac {q^{2}m\omega }{\hbar }).
\label{v3}
\end{equation}
The Husimi Q--function equals
\begin{eqnarray}
Q_{0}(p,q)&=&<\beta |0><0|\beta >
=\exp (-\frac {p^{2}}{2\hbar m \omega }
-\frac {q^{2}m\omega }{2\hbar }),\nonumber\\
\beta &=&\frac {1}{\sqrt 2}(q\sqrt {\frac {m\omega }{\hbar }}+
i\frac {p}{\sqrt {\hbar m\omega }}).
\label{v4}
\end{eqnarray}
Since Q--function is the diagonal matrix element of the density
operator in coherent state basis it is easy to obtain the Q--function
of the thermal equilibrium state of the oscillator at temperature $~T$
with density operator
\begin{equation}
\rho (T)=2\sinh \frac {\hbar \omega }{2T}\exp [-\frac {\hbar \omega }{T}
(a\dag a+\frac {1}{2})].
\label{v5}
\end{equation}
The operator $~Z\rho (T)$, where $~Z$ is the partition function
\begin{equation}
Z=(2\sinh \frac {\hbar \omega }{2T})^{-1},
\label{v6}
\end{equation}
coincides with the evolution operator $~U(t)$ if one takes
$~it=~\hbar /T$. This equality gives in coherent state basis the
Q--function of the oscillator at temperature $~T$
\begin{equation}
Q_{T}(q,p)=2\sinh \frac {\hbar \omega }{2T}\exp \{-\frac {\hbar \omega }
{2T}-\frac {1}{2}(\frac {p^{2}}{\hbar \omega m}+\frac {q^{2}m\omega }
{\hbar })[1-\exp (-\frac {\hbar \omega }{T})]\},
\label{v7}
\end{equation}
if one uses formula (\ref{111}) for the oscillator propagator in
coherent state basis
$$|\beta >=|q\sqrt {\frac {m\omega }{2\hbar }}
+i\frac {p}{\sqrt {2\hbar m\omega }}>.$$

\section{Multimode Mixed Correlated Light}

The most general mixed squeezed state of the N--mode light with a
Gaussian density operator $~\hat{\rho }$ is described by
the Wigner function $~W({\bf p},{\bf q})$ of the generic Gaussian form,
\begin{equation}
W({\bf p},{\bf q})=(\det {\bf M})^{-\frac 12}\exp\left
[-\frac 12({\bf Q}-<{\bf Q}>){\bf M}^{-1}({\bf Q}-<{\bf Q}>)\right],
\label{m1}
\end{equation}
where 2N--dimensional vector $~{\bf Q}=({\bf p},{\bf q})$ consists
of N components $~p_1,~\ldots ,~p_N$ and N components
$~q_1,~\ldots ,~q_N$;
operators $~\hat {{\bf p}}$ and $~\hat {{\bf q}}$ being
quadrature components of the photon creation
$~\hat {{\bf a}}\dag$ and annihilation $~\hat {{\bf a}}$
operators (we use dimensionless variables and assume $~\hbar =1$):
\begin{eqnarray}
\hat {{\bf p}}&=&\frac {\hat {{\bf a}}
-\hat {{\bf a}}\dag}{i\sqrt {2}},\nonumber\\
\hat {{\bf q}}&=&\frac {\hat {{\bf a}}
+\hat {{\bf a}}\dag}{\sqrt {2}}.
\label{m2}
\end{eqnarray}
2N parameters $~<p_i>$ and $~<q_i>$, $~i=1,2,\ldots ,N$, combined
into vector $~<{\bf Q}{\bf >}$, are average values of
quadratures,
\begin{eqnarray}
<{\bf p}>&=&\mbox{Tr}~\hat{\rho}\hat {{\bf p}},\nonumber\\
<{\bf q}>&=&\mbox{Tr}~\hat{\rho}\hat {{\bf q}}.
\label{m3}
\end{eqnarray}
A real symmetric dispersion matrix $~{\bf M}$ consists of 2N$^2$+N
variances
\begin{equation}
{\cal M}_{\alpha\beta}=\frac 12\left\langle\hat Q_{\alpha}\hat
Q_{\beta}+\hat Q_{\beta}\hat Q_{\alpha}\right\rangle -\left\langle
\hat Q_{\alpha}\right\rangle\left\langle\hat Q_{\beta}\right\rangle
,~~~~~~~~~~~\alpha ,\beta =1,2,\ldots ,2N.
\label{m4}
\end{equation}
They obey certain constraints, which are nothing but
generalized uncertainty relations \cite{dod89}.

The photon distribution function of this state
\begin{equation}
{\cal P}_{{\bf n}}=\mbox{Tr}~\hat{\rho }|{\bf n}><{\bf n}|,
{}~~~~~~~{\bf n}=(n_1,n_2,\ldots ,n_N),
\label{m5}
\end{equation}
where the state $~|{\bf n}>$ is photon number state, which was
calculated in \cite{olga}, \cite{semen} and it is
\begin{equation}
{\cal P}_{{\bf n}}={\cal P}_0\frac {H_{{\bf n}{\bf n}}^{
\{{\bf R}\}}({\bf y})}{{\bf n}!}.
\label{m6}
\end{equation}
The function $~H_{{\bf n}{\bf n}}^{\{{\bf R}\}}({\bf y})$ is
multidimensional Hermite polynomial. The probability to have no
photons is
\begin{equation}
{\cal P}_0=\left[\det\left({\bf M}+\frac 12{\bf I}_{2N}\right)\right
]^{-\frac 12}\exp\left[-<{\bf Q}>\left(2{\bf M}+{\bf I}_{2N}\right
)^{-1}<{\bf Q}>\right],
\label{m7}
\end{equation}
where we introduced the matrix
\begin{equation}
{\bf R}=2{\bf U}^{\dag }(1+2{\bf M})^{-1}{\bf U}^{*}-\sigma _{Nx},
\label{m8}
\end{equation}
and the matrix
\begin{equation}
\sigma _{Nx}=\left(\begin{array}{cc}
0&{\bf I}_N\\
{\bf I}_N&0\end{array}\right).
\label{m9}
\end{equation}
The argument of Hermite polynomial is
\begin{equation}
{\bf y}=2{\bf U}^t({\bf I}_{2N}-2{\bf M})^{-1}<{\bf Q}>,
\label{m10}
\end{equation}
and the 2N--dimensional unitary matrix
\begin{equation}
{\bf U}=\frac 1{\sqrt {2}}\left(\begin{array}{cc}
-i{\bf I}_N&i{\bf I}_N\\
{\bf I}_N&{\bf I}_N\end{array}\right)
\label{m11}
\end{equation}
is introduced, in which $~{\bf I}_N$ is the N$\times $N identity
matrix. Also we use the notation
$${\bf n}!=n_1!n_2!\cdots n_N!.$$

The mean photon number for j--th mode is expressed in terms of
photon quadrature means and dispersions
\begin{eqnarray}
<n_j>=\frac 12(\sigma_{p_jp_j}+\sigma_{q_jq_j}-1)
+\frac 12(<p_j>^2+<q_j^2>).
\label{m12}
\end{eqnarray}
We introduce a complex 2N--vector $~{\bf B}=(\beta _{1},~\beta _{2},
{}~\ldots ,~\beta _{N},~\beta _{1}^{*},~\beta _{2}^{*},~\ldots ,
{}~\beta _{N}^{*})$. Then the Q--function \cite{hus40} is
the diagonal matrix element of the density operator in coherent
state basis $~|~\beta _{1},~\beta _{2},~\ldots ,~\beta _{N}>.$
This function is the generating function for matrix elements of the
density operator in the Fock basis $~|${\bf n}$>$ which has been calculated
in \cite{semen}. In notations corresponding to the Wigner function
(\ref{m1}) the Q--function is
\begin{equation}
Q({\bf B})={\cal P}_{0}\exp [-\frac {1}{2}{\bf B}(R+\sigma _{Nx}){\bf B}+
{\bf B}R{\bf y}].
\label{m13}
\end{equation}
Thus, if the Wigner function (\ref{m1}) is given one has the Q--function.
Also, if one has the Q--function (\ref{m13}), i.e. the matrix $~R$ and
the vector {\bf y}, the Wigner function may be obtained due to
relations
\begin{eqnarray}
{\bf M}&=&{\bf U}^{*}(R+\sigma _{Nx})^{-1}{\bf U}^{\dag }
-\frac {1}{2},\nonumber\\
<{\bf Q}>&=&{\bf U}^{*}[1-(R+\sigma _{Nx})^{-1}\sigma _{Nx}]{\bf y}.
\label{m14}
\end{eqnarray}
For pure squeezed and correlated state with the wave function
\begin{equation}
\Psi =N\exp [-{\bf x}m{\bf x}+{\bf c}{\bf x}],
\label{m15}
\end{equation}
where
\begin{equation}
N=[\det (m+m^{*})]^{1/4}\pi ^{-N/4}\exp \{\frac {1}{4}({\bf c}
+{\bf c}^{*})(m+m^{*})^{-1}({\bf c}+{\bf c}^{*})\},
\label{m16}
\end{equation}
the symmetric 2N$\times $2N--matrix $~R$ determining Q--function
has the block--diagonal form
\begin{equation}
R=\left( \begin{array}{clcr}r&0\\
0&r^{*}\end{array}\right).
\label{m17}
\end{equation}
The N$\times $N--matrix $~r$ is expressed in terms of symmetric matrix
$~m$
\begin{equation}
r^{*}=1-(m+1/2)^{-1},
\label{m18}
\end{equation}
and the 2N--vector $~{\bf y}=({\bf Y,Y}^{*})$ is given by the
relation
\begin{equation}
{\bf Y}^{*}=\frac {1}{\sqrt 2}(m-1/2)^{-1}{\bf c}.
\label{m19}
\end{equation}
Corresponding blocks of the dispersion matrix
\begin{equation}
{\bf M}=\left( \begin{array}{clcr}\sigma _{\bf pp}&\sigma _{\bf pq}\\
\widetilde \sigma _{\bf pq}&\sigma _{\bf qq}\end{array}\right)
\label{m20}
\end{equation}
are
\begin{eqnarray}
\sigma _{\bf pp}&=&2(m^{-1}+m^{*-1})^{-1},\nonumber\\
\sigma _{\bf qq}&=&\frac {1}{2}(m+m^{*})^{-1},\nonumber\\
\sigma _{\bf pq}&=&\frac {i}{2}(m-m^{*})(m+m^{*})^{-1}.
\label{m21}
\end{eqnarray}
The probability to have no photons is
$$
P_{0}=\frac {[\det (m+m^{*})]^{1/2}}{|\det (m+1/2)|}$$
\begin{equation}
\otimes \exp \{\frac {1}{2}({\bf c}+{\bf c}^{*})(m+m^{*})^{-1}({\bf c}
+{\bf c}^{*})+\frac {1}{4}[{\bf c}(m+1/2)^{-1}{\bf c}
+{\bf c}^{*}(m^{*}+1/2)^{-1}{\bf c}^{*}]\}.
\label{m22}
\end{equation}
Multivariable Hermite polynomials describe the photon distribution
function for the multimode mixed and pure correlated light \cite{olga},
\cite{md94}, \cite{dodon94}. The nonclassical state of light may be
created due to nonstationary Casimir effect \cite{jslr}, and the
multimode oscillator is the model to describe the behaviour of
squeezed and correlated photons.

\section{Multivariable Hermite Polynomials}

For parametric forced oscillator, the transition amplitude between its
energy levels has been calculated as overlap integral of two generic
Hermite polynomials with a Gaussian function (Frank--Condon factor)
and expressed in terms of Hermite polynomials of two variables
\cite{mal70}. For N--mode parametric oscillator, the analogous amplitude
has been expressed in terms of Hermite polynomials of 2N variables, i.e.
the overlap integral of two generic Hermite polynomials of N variables
with a Gaussian function (Frank--Condon factor for a polyatomic molecule)
has been evaluated in \cite{MalManTri73}, \cite{tri73}. The corresponding
result uses the formula
$$({\bf n}=n_{1},n_{2},\ldots ,n_{N},~~~~~{\bf m}=m_{1},m_{2},\ldots
m_{N},~~~~~m_{i},n_{i}=0,1,\ldots)$$
\begin{equation}
\int H_{{\bf n}}^{\{R\}}({\bf x})H_{{\bf m}}^{\{r\}}(\Lambda {\bf x}
+{\bf d})\exp (-{\bf x}m{\bf x}+{\bf c}{\bf x})d{\bf x}
=\frac {\pi ^{N/2}}{\sqrt {\det m}}\exp (\frac {1}{4}{\bf c}
m^{-1}{\bf c})H_{{\bf mn}}^{\{\rho \}}({\bf y}),
\label{p52}
\end{equation}
where the symmetric 2N$\times $2N--matrix
\begin{equation}
\rho=\left( \begin{array}{clcr}R_{1}&R_{12}\\
\widetilde R_{12}&R_{2}\end{array}\right)
\label{p53}
\end{equation}
with N$\times $N--blocks $~R_{1},~R_{2},~R_{12}$ is expressed in terms
of symmetric N$\times $N--matrices $~R,~r,~m$ and N$\times $N--matrix
$~\Lambda $ in the form
\begin{eqnarray}
R_{1}&=&R-\frac {1}{2}Rm^{-1}R,\nonumber\\
R_{2}&=&r-\frac {1}{2}r\Lambda m^{-1}\widetilde \Lambda r,\nonumber\\
\widetilde R_{12}&=&-\frac {1}{2}r\Lambda m^{-1}R.
\label{p54}
\end{eqnarray}
Here the matrix $~\widetilde \Lambda $ is transposed matrix $~\Lambda $,
and $~\widetilde R_{12}$ is transposed matrix $~R_{12}.$ The 2N--vector
$~{\bf y}$ is expressed in terms of N--vectors $~{\bf c}$ and $~{\bf d}$
in the form
\begin{equation}
{\bf y}=\rho ^{-1}\left( \begin{array}{c}{\bf y}_{1}\\
{\bf y}_{2}\end{array}\right ),
\label{p55}
\end{equation}
where N--vectors $~{\bf y}_{1}$ and $~{\bf y}_{2}$ are
\begin{eqnarray}
{\bf y}_{1}&=&\frac {1}{4}(Rm^{-1}+m^{-1}R){\bf c}\nonumber\\
{\bf y}_{2}&=&\frac {1}{4}(r\Lambda m^{-1}
+m^{-1}\widetilde \Lambda r){\bf c}+r{\bf d}.
\label{p56}
\end{eqnarray}
For matrices $~R=2,~r=2,$ the above formula (\ref{p52}) yields
$$\int \{\prod _{i=1}^{N}H_{n_{i}}(x_{i})H_{m_{i}}
(\sum _{k=1}^{N}\Lambda _{ik}x_{k}+d_{i})\}\exp (-{\bf x}m{\bf x}
+{\bf c}{\bf x})d{\bf x}$$
\begin{equation}
=\frac {\pi ^{N/2}}{\sqrt {\det m}}
\exp (\frac {1}{4}{\bf c}m^{-1}{\bf c})H_{{\bf mn}}^{\{\rho \}}({\bf y}),
\label{p57}
\end{equation}
with N$\times $N--blocks $~R_{1},~R_{2},~R_{12}$ expressed in terms of
N$\times $N--matrices $~m$ and $~\Lambda $
in the form
\begin{eqnarray}
R_{1}&=&2(1-m^{-1}),\nonumber\\
R_{2}&=&2(1-\Lambda m^{-1}\widetilde \Lambda ),\nonumber\\
\widetilde R_{12}&=&-2\Lambda m^{-1}.
\label{p58}
\end{eqnarray}
The 2N--vector
$~{\bf y}$ is expressed in terms of N--vectors $~{\bf c}$ and
$~{\bf d}$ in the form (\ref{p55}) with
\begin{eqnarray}
{\bf y}_{1}&=&m^{-1}{\bf c},\nonumber\\
{\bf y}_{2}&=&\frac {1}{2}(\Lambda m^{-1}
+m^{-1}\widetilde \Lambda ){\bf c}+2{\bf d}.
\label{p59}
\end{eqnarray}
If the symmetric matrix $~\rho $ has the block--diagonal structure
$$
\rho=\left( \begin{array}{clcr}R_{1}&0\\
0&R_{2}\end{array}\right),
$$
with the symmetric S$\times $S--matrix $~R_{1}$ and the symmetric
(2N-S)$\times $(2N-S)--matrix $~R_{2}$, the multivariable Hermite
polynomial is represented as the product of two Hermite polynomials
depending on S and 2N-S variables, respectively,
$$
H_{\bf k}^{\{{\bf \rho }\}}({\bf y})
=H_{{\bf n}_{S}}^{\{{\bf R}_{1}\}}({\bf y}_{1})
H_{{\bf n}_{2N-S}}^{\{{\bf R}_{2}\}}({\bf y}_{2}),
$$
where the 2N--vector $~{\bf y}$ has vector--components
$${\bf y}=(~{\bf y}_{1},~{\bf y}_{2}),$$
and the 2N--vector $~{\bf k}$ has components
$${\bf k}=({\bf n}_{S},~{\bf n}_{2N-S})=(~n_{1},~\ldots ,~n_{S},~n_{S
+1},~\ldots ,~n_{2N}).$$
The partial case of this relation is the relation for Hermite
polynomials with the matrix $~R$ with complex conjugate blocks
$~R_{1}=~r,~~R_{2}=~r^{*}$, and complex conjugate vector--components
$~{\bf y}_{1}=~{\bf y}_{2}^{*}$
$$
H_{\bf k}^{\{{\bf \rho }\}}({\bf y})
=|H_{{\bf n}_{S}}^{\{{\bf R}_{1}\}}({\bf y}_{1})|^{2},~~~S=N.$$
The calculated integrals are important to evaluate the Green function
or density matrix for systems with quadratic Hamiltonians.
Partial cases of multivariable Hermite polynomials determine some other
special functions \cite{md94}, \cite{dodon94}.

\section{Multimode Even and Odd Coherent States}

We define multimode even and odd coherent states (Schr\"odinger
cat male states and Schr\"odinger cat female states, respectively)
as \cite{ans}
\begin{equation}
\mid {\bf A_{\pm}}>=N_{\pm} (\mid {\bf A}> \pm \mid -{\bf A}>),
\label{e23}
\end{equation}
where the multimode coherent state $\mid {\bf A}>$ is
\begin{equation}
\mid {\bf A}>=\mid \alpha_{1},~\alpha_{2},~\ldots ,~\alpha_{n}>
=D({\bf A}) \mid {\bf 0}>,
\label{e24}
\end{equation}
and the multimode coherent state is created from the multimode vacuum
state $\mid {\bf 0}>$ by the multimode displacement operator $D({\bf A})$.
The definition of multimode even and odd coherent states is the obvious
generalization of the one--mode even and odd coherent state given in
\cite{mal79}, \cite{dod74}.
Normalization constants for multimode even and odd coherent states
become
\begin{eqnarray}
N_{+}&=&\frac{e^{\mid {\bf A} \mid^{2}/2}}
{2 \sqrt{\cosh\mid {\bf A}\mid^{2}}},\nonumber\\
N_{-}&=&\frac{e^{\mid {\bf A} \mid^{2}/2}}
{2 \sqrt{\sinh\mid {\bf A}\mid^{2}}},
\label{e25}
\end{eqnarray}
where $~{\bf A}=(~\alpha_{1},~\alpha_{2},~\ldots,~\alpha_{n})$ is a
complex vector and its modulus is
\begin{equation}
\mid {\bf A} \mid^{2}=\sum_{m=1}^{n}\mid \alpha_{m} \mid^{2}.
\label{e26}
\end{equation}
Such multimode even and odd coherent states can be decomposed into
multimode number states as
\begin{equation}
\mid {\bf A_{\pm}}>=N_{\pm}\sum_{{\bf n}}
\frac{e^{-\mid {\bf A}\mid^{2}/2}
\alpha_{1}^{n_{1}}\cdots \alpha_{n}^{n_{n}}}{\sqrt{n_{1}!}
\cdots \sqrt{n_{n}!}}
(1\pm(-1)^{n_{1}+n_{2}+\cdots n_{n}}) \mid {\bf n}>,
\label{e27}
\end{equation}
where $\mid {\bf n}>=~\mid n_{1},~n_{2},~\ldots ,~n_{n}>$ is
multimode number state. Also from Eq. (\ref{e23}), we can derive an
important relation for the multimode even and odd coherent states,
namely,
\begin{eqnarray}
a_{i} \mid {\bf A_{+}}>&=&\alpha_{i} \sqrt{\tanh\mid {\bf A}\mid^{2}}
\mid {\bf A_{-}}>,\nonumber\\
a_{i} \mid {\bf A_{-}}>&=&\alpha_{i} \sqrt{\coth\mid {\bf A}\mid^{2}}
\mid {\bf A_{+}}>.
\label{e28}
\end{eqnarray}
The probability of finding {\bf n} photons in multimode even and odd
coherent states can be worked out with the help of Eq. (\ref{e27})
\begin{eqnarray}
P_{+} ({\bf n})&=&\frac{\mid \alpha_{1} \mid^{2n_{1}}
\mid \alpha_{2} \mid^{2n_{2}}\cdots \mid \alpha_{n} \mid^{2n_{n}} }
{(n_{1}!)(n_{2}!)\cdots (n_{n}!)\cosh\mid {\bf A}\mid^{2}},
{}~~~n_{1}+n_{2}+\cdots +n_{n}=2k,\nonumber\\
 P_{-} ({\bf n})&=&\frac{\mid \alpha_{1} \mid^{2n_{1}}
\mid \alpha_{2} \mid^{2n_{2}}\cdots \mid \alpha_{n} \mid^{2n_{n}} }
{(n_{1}!)(n_{2}!)\cdots (n_{n}!)\sinh\mid {\bf A}\mid^{2}},
{}~~~n_{1}+n_{2}+\cdots +n_{n}=2k+1.
\label{e29}
\end{eqnarray}
Multimode coherent states are the product of independent coherent states
of each mode, and photon distribution function is the product of
independent Poissonian distribution functions. But in the present case
of multimode even and odd coherent states we cannot factorize their
multimode photon distribution functions due to the presence of the
nonfactorizable $\cosh\mid {\bf A}\mid^{2}$ and
$\sinh\mid {\bf A}\mid^{2}$. This fact implies the phenomenon of
statistical dependences of different modes of these states on each other.

In order to describe properties of distribution functions from
Eq. (\ref{e29}) we will calculate the symmetric 2N$\times $2N
dispersion matrix for multimode field quadrature components. For even
and odd coherent states, we have
\begin{equation}
<{\bf A}_{\pm} \mid a_{i}a_{k} \mid {\bf A}_{\pm}>=\alpha_{i}\alpha_{k},
\label{e30}
\end{equation}
and complex conjugate values of the above equation for
$<{\bf A}_{\pm} \mid a_{i}^{\dag}a_{k}^{\dag} \mid {\bf A}_{\pm}>$.
Since the quantity $<{\bf A}_{\pm} \mid a_{i} \mid {\bf A}_{\pm}>$ is
equal to zero the above equation gives two N$\times $N blocks of the
dispersion matrix. For other two N$\times $N blocks of this matrix,
we have for multimode even coherent states
\begin{equation}
\sigma_{(a_{i}^{\dag}a_{k})}^{+}
=<{\bf A}_{+} \mid \frac{1}{2}(a_{i}^{\dag}a_{k}
+a_{k}a_{i}^{\dag})\mid {\bf A_{+}}>=\alpha_{i}^{*}\alpha_{k}
\tanh\mid {\bf A} \mid^{2}+\frac{1}{2} \delta_{ik},
\label{e31}
\end{equation}
and for multimode odd coherent states
\begin{equation}
\sigma_{(a_{i}^{\dag}a_{k})}^{-}
=<{\bf A}_{-} \mid \frac{1}{2}(a_{i}^{\dag}a_{k}
+a_{k}a_{i}^{\dag})\mid {\bf A_{-}}>=\alpha_{i}^{*}\alpha_{k}
\coth\mid {\bf A} \mid^{2}+\frac{1}{2} \delta_{ik}.
\label{e32}
\end{equation}
For the dispersion matrix, the mean values of the photon
numbers $~n_{i}=a_{i}^{\dag}a_{i}$ for multimode even and odd
coherent states are the following
\begin{eqnarray}
<{\bf A_{+}}\mid n_{i} \mid {\bf A_{+}}>&=&\mid \alpha_{i}\mid^{2}
\tanh\mid {\bf A}\mid^{2},\nonumber\\
<{\bf A_{-}}\mid n_{i} \mid {\bf A_{-}}>&=&\mid \alpha_{i}\mid^{2}
\coth\mid {\bf A}\mid^{2}.
\label{e33}
\end{eqnarray}
Taking into account the above equation the symmetric N$\times $N
dispersion matrices for photon number operators can be obtained from
the above given distribution functions for multimode even and odd
coherent states. By defining
\begin{equation}
\sigma_{ik}^{\pm}=<{\bf A}_{\pm}\mid n_{i}n_{k}\mid
{\bf A}_{\pm}>,
\label{e34}
\end{equation}
corresponding  expressions in such states are
\begin{eqnarray}
\sigma _{ik}^{+}&=
&\mid \alpha _{i} \mid ^{2}\mid \alpha _{k}\mid ^{2}
\mbox {sech }^{2}\mid {\bf A}\mid ^{2}+\mid \alpha _{i}\mid ^{2}
\tanh \mid {\bf A}\mid ^{2}\delta _{ik},\nonumber\\
\sigma _{ik}^{-}&=&-\mid \alpha _{i}\mid ^{2}\mid \alpha _{k}\mid^{2}
\mbox {csch }^{2}\mid {\bf A}\mid ^{2}+\mid \alpha _{i}\mid ^{2}
\coth \mid {\bf A}\mid ^{2}\delta _{ik}.
\label{e35}
\end{eqnarray}
As nondiagonal matrix elements of the dispersion density matrix
are not equal to zero, we can predict that different modes of
these states are correlated with each other. In other words, as we have
mentioned before, there exist some statistical dependences of
different modes on each other.

Another interesting property for multimode even and odd coherent
states is the Q--function, and it can be obtained in the following
manner. First of all the density matrices for multimode even and
odd coherent states are
\begin{equation}
\rho_{\pm}=\mid {\bf A_{\pm}}><{\bf A_{\pm}} \mid,
\label{e36}
\end{equation}
then the Q--function can be calculated as
\begin{eqnarray}
 Q_{+}({\bf B},{\bf B}^{*})&=&<{\bf B} \mid \rho_{+}
\mid {\bf B}>\nonumber\\
&=&4N_{+}^{2}e^{-(\mid {\bf A}\mid ^{2}
+\mid {\bf B}\mid ^{2})}\mid
\cosh({\bf A {\bf B}^{*}})\mid ^{2},\nonumber\\
Q_{-}({\bf B},{\bf B}^{*})&=&<{\bf B} \mid \rho_{-}
\mid {\bf B}>\nonumber\\
&=&4N_{-}^{2}e^{-(\mid {\bf A}\mid ^{2}
+\mid {\bf B}\mid ^{2})}\mid \sinh({\bf A
{\bf B}^{*}})\mid ^{2},
\label{e37}
\end{eqnarray}
where $\mid {\bf B}>=\mid \beta_{1},~\beta_{2},~\ldots ,~\beta_{n}>$
is another multimode coherent state with multimode eigenvalue
${\bf B}=(\beta_{1},~\beta_{2},~\ldots ,~\beta_{n})$.
We call these functions for even and odd coherent states
as the Q--functions for Schr\"odinger cat states.
The Q--function for single--mode odd coherent state shows the
crater type behaviour for small values of the quantity
$\mid \alpha \mid$ and for its larger values the Q--function begins
to split into two peaks in a similar manner as in case of even coherent
states \cite{ans}.

The Wigner function for multimode coherent states is \cite{dod89}
\begin{equation}
W_{{\bf A,B}}=2^{N}\exp[-2{\bf Z Z^{*}}+2{\bf A Z^{*}}
+2{\bf B^{*}Z}-{\bf AB^{*}}-\frac{\mid {\bf A}\mid^{2}}{2}
-\frac{\mid {\bf B}\mid^{2}}{2}],
\label{e38}
\end{equation}
where
\begin{equation}
{\bf Z}=\frac{{\bf q+i p}}{\sqrt{2}}.
\label{e39}
\end{equation}
For even and odd coherent states, the Wigner  function is
\begin{eqnarray}
W_{{\bf A}_{\pm}}({\bf q,p})
&=&\mid N_{\pm}\mid^{2}[W_{{\bf (A,B=A)}}({\bf q,p})\pm
W_{{\bf (A,B=-A)}}({\bf q,p})\nonumber\\
&\pm &W_{{\bf (-A,B=A)}}({\bf q,p})+W_{({\bf -A,B=-A)}}({\bf q,p})],
\label{e40}
\end{eqnarray}
where explicit forms of $N_{\pm}$ are given in Eq. (\ref{e25}).
For multimode case, we use following notations
\begin{eqnarray}
{\bf AZ^{*}}&=&\alpha_{1}Z_{1}^{*}+\alpha_{2}Z_{2}^{*}
+\cdots \alpha_{n}Z_{n}^{*},
\nonumber\\
{\bf ZZ^{*}}&=&Z_{1}Z_{1}^{*}+Z_{2}Z_{2}^{*}+\cdots +Z_{n}Z_{n}^{*}.
\label{e41}
\end{eqnarray}

The photon distribution function gives the probability of finding
$~2k$ photons for two--mode even coherent state, and is defined as
\begin{equation}
P_{+}(2k)=\frac{(\mid \alpha_{1} \mid^{2}
+\mid \alpha_{2} \mid^{2})^{2k}}
{(2k)!\cosh(\mid \alpha_{1} \mid^{2}+\mid \alpha_{2} \mid^{2})},
\label{e42}
\end{equation}
where $2k=n_{1}+n_{2}$, for both $n_{1}$ and $n_{2}$ to be even or
odd numbers. Similarly for two--mode odd coherent state it gives the
probability of finding $~2k+1$ photons
\begin{equation}
P_{-}(2k+1)=\frac{(\mid \alpha_{1} \mid^{2}
+\mid \alpha_{2} \mid^{2})^{(2k+1)}}
{(2k+1)!\sinh(\mid \alpha_{1} \mid^{2}+\mid \alpha_{2} \mid^{2})}.
\label{e43}
\end{equation}
For this case, $~n_{1}$ is even and $~n_{2}$ is odd number, or vice versa.
For single--mode case, the photon distribution functions demonstrate
super and sub--Poissonian properties for even and odd coherent states,
respectively. The same conclusion may be drawn for two--mode
(and multimode) even and odd coherent states.

\section{Generic Quadratic Systems}

The generic mechanical or optical linear system with energy operator
has the Hamiltonian
\begin{equation}\label{gq1}
H=\frac {1}{2}{\bf Q}B(t){\bf Q}+{\bf C}(t){\bf Q},
\end{equation}
where we use 2N--vectors $~{\bf Q}=~(p_{1},~p_{2},\ldots ,
p_{N},~q_{1},~q_{2},\ldots ,q_{N}),$ and $~{\bf C}(t)~$ as
well as, 2N$\times $2N--matrix $~B(t)$, the Planck constant
$~\hbar =1.$ This system has 2N linear integrals of motion
\cite{dod89}, \cite{mal79} which are written in vector form
\begin{equation}\label{gq2}
{\bf Q}_{0}(t)=\Lambda (t){\bf Q}+{\bf \Delta }
(t).
\end{equation}
The real symplectic matrix $~\Lambda (t)~$ is the solution of
the system of equations
\begin{eqnarray}\label{gq3}
\dot \Lambda (t)&=&\Lambda (t)\Sigma B(t),\nonumber\\
\Lambda (0)&=&1,
\end{eqnarray}
where the real antisymmetrical matrix  $~\Sigma ~$ is 2N--dimensional
analog of the Pauli matrix $~i\sigma _{y},$ and the vector
$~{\bf \Delta }(t)~$ is solution of the system of equations
\begin{eqnarray}\label{gq4}
\dot {\bf \Delta }(t)&=&\Lambda (t)\Sigma {\bf C}(t),\nonumber\\
{\bf \Delta }(0)&=&0.
\end{eqnarray}
If for time $~t=~0,$ one has the initial Wigner function of the system
in the form
\begin{equation}\label{gq5}
W({\bf p},{\bf q},t=0)=W_{0}({\bf Q}),
\end{equation}
the Wigner function of the system at time $~t~$ is
\begin{equation}\label{gq6}
W({\bf p},{\bf q},t)=W_{0}[\Lambda (t){\bf Q}+{\bf \Delta }(t)].
\end{equation}
The Hamiltonian (\ref{gq1}) may be rewritten in terms of creation and
annihilation operators
\begin{equation}\label{gq7}
H=\frac {1}{2}{\bf A}D(t){\bf A}
+{\bf E}(t){\bf A},
\end{equation}
where we use 2N--vectors $~{\bf A}=~(~a_{1},~a_{2},\ldots ,
{}~a_{N},~a_{1}\dag ,~a_{2}\dag ,\ldots ,~a_{N}\dag ),$ and
$~{\bf E}(t)~$ as well as, 2N$\times $2N--matrix $~D(t).$
This system has 2N--linear integrals of motion \cite{dod89},
\cite{mal79} which are written in vector form
\begin{equation}\label{gq8}
{\bf A}_{0}(t)=M(t){\bf A}+{\bf N}(t).
\end{equation}
The complex matrix $~M(t)~$ is the solution of the system of
equations
\begin{eqnarray}\label{gq9}
\dot M(t)&=&M(t)\sigma D(t),\nonumber\\
M(0)&=&1,
\end{eqnarray}
where the imaginary antisymmetric matrix $~\sigma ~$ is
2N$\times $2N--analog of the Pauli matrix $~-\sigma _{y},$ and
the vector $~{\bf N}(t)~$ is solution of the system of equations
\begin{eqnarray}\label{gq10}
\dot {\bf N}(t)&=&M(t)\sigma {\bf E}(t),\nonumber\\
{\bf N}(0)&=&0.
\end{eqnarray}
If for time $~t=~0,$ one has the initial Q--function of the system
in the form
\begin{equation}\label{gq11}
Q({\bf \alpha },{\bf \alpha }^{*},t=0)
=Q_{0}({\bf \alpha }),
\end{equation}
the Q--function of the system at time $~t~$ is
\begin{equation}\label{gq12}
Q({\bf \alpha },{\bf \alpha }^{*},t)=
Q_{0}[M(t){\bf \alpha }+{\bf N}(t)].
\end{equation}
Here $~{\bf \alpha }=~(~{\bf q}~+~i{\bf p})/\sqrt 2.$

For time--independent Hamiltonian (\ref{gq1}), the matrix
$~\Lambda (t)~$ is
\begin{equation}\label{gq13}
\Lambda (t)=\exp (\Sigma Bt),
\end{equation}
and the vector $~{\bf \Delta }(t)~$ is
\begin{equation}\label{gq14}
{\bf \Delta }(t)
=\int _{0}^{t}\exp (\Sigma B\tau )~\Sigma ~{\bf C}(\tau )~d\tau .
\end{equation}
For time--independent Hamiltonian (\ref{gq7}), the matrix $~M(t)~$ is
\begin{equation}\label{gq15}
M(t)=\exp (\sigma Dt),
\end{equation}
and the vector $~{\bf N}(t)~$ is
\begin{equation}\label{gq16}
{\bf N}(t)
=\int _{0}^{t}\exp (\sigma D\tau )~\sigma ~{\bf E}(\tau )~d\tau .
\end{equation}
For time--dependent linear systems, the Wigner function of generic squeezed
and correlated state has Gaussian form and it was calculated in \cite{dod89}.

\section{Optical Homodyne Tomography}

In \cite{vogel} it was shown a relation of marginal distribution
$~w(X,\Theta )~$ for homodyne output variable
\begin{equation}\label{ht1}
\hat X(\Theta )=\hat q\cos \Theta -\hat p\sin \Theta ,
\end{equation}
and the Wigner function $~W(q,p)~$  of the form
\begin{equation}\label{ht2}
w(X,\Theta )=\int W(x\cos \Theta -v\sin \Theta ,
x\sin \Theta +v\cos \Theta )~dv,
\end{equation}
where
\begin{equation}\label{ht3}
\int _{-\infty }^{\infty }w(X,\Theta )~dX=1.
\end{equation}
The inverse relation which is Radon tomographic transform,
\begin{eqnarray}\label{ht4}
W(q,p,s\rightarrow 0)&=&\frac {1}{4\pi ^{2}}\int _{-\infty }^{\infty }
dx~dy\int_{0}^{\pi }d\Theta ~w(X,\Theta )\nonumber\\
&\times &\exp [(sy^{2}/8)+iy(X-q\cos \Theta -p\sin \Theta )]~|y|,
\end{eqnarray}
gives the possibility measuring the homodyne variables (\ref{ht1}) to
measure the Wigner function of the system. In \cite{tombesi} the method
was extended to measure the Wigner function by measuring the marginal
distribution $~w(X,\mu ,\nu ,\delta )~$ for the variable
\begin{equation}\label{ht5}
\hat X=\mu \hat q+\nu \hat p +\delta ,
\end{equation}
since it is easy to show that the Wigner function is
\begin{equation}\label{ht6}
W(q,p)=(2\pi )^{2}s^{2}\exp (isX)~\widetilde w(X,sq,sp,s).
\end{equation}
The function $~\widetilde w(X,m,n,s)~$ is the Fourier transform of the
marginal distribution of $~w(X,\mu ,\nu ,\delta ).$

\section{Conclusion}

The discussed method of the time--dependent integrals of motion may be
applied to systems with varying parameters, for example, to the model of
the creation of photons and squeezing phenomenon in frame of nonstationary
Casimir effect using the generalized harmonic oscillator (see,
\cite{casta}).
The analogous phenomenon of reduction of quantum noise for oscillating
mirrors has been discussed in \cite{tombesiman}. The behaviour of the
system containing moving mirrors and light beams has been discussed in
\cite{solimeno}, and using even and odd coherent state light
(Schr\"odinger cat light) in interferometric gravitational wave detectors
has been suggested in \cite{solzacman}.

The coherent states \cite{gla63}, \cite{sud}, \cite{kla} which describe
classical states of light have the Poisson photon statistics.
Nonclassical states like squeezed states \cite{holl}, \cite{yuen},
\cite{agar94} or generalized correlated states \cite{sudar} have
nonpoissonian light statistics. Different kinds of nonclassical states
are studied in \cite{nieto}.

There is possibility to introduce different types of nonclassical states
using different types of superposition of complete set either of photon
number states or of coherent states. All such states which may be
constructed experimentally are interesting, and many of these states may
be created in situations with nonstationarities or in situations
with nonlinearities.

\section*{Acknowledgments}
The author would like to acknowledge the
Organizers of Latin American School on Physics 1995
for kind hospitality.

\end{document}